\documentclass[12pt]{article}
\usepackage{array}
\usepackage{graphicx}
\usepackage{amssymb}
\usepackage{amsmath}
\input{epsf}
\usepackage{cite}
\def\@fmsl@sh#1#2#3{\m@th\ooalign{$\hfil#1\mkern#2/\hfil$\crcr$#1#3$}}
 \def\eq#1\en{\begin{equation}#1\end{equation}}
\def\s[#1,#2]{[#1\stackrel{\star}{,}#2]}
\def\sx[#1,#2]{[#1\stackrel{\star_{x}}{,}#2]}

\newcommand{\nc}{\newcommand}
\nc{\beq}{\begin{equation}}
\nc{\eeq}{\end{equation}}
\nc{\beqa}{\begin{eqnarray}}
\nc{\eeqa}{\end{eqnarray}}

\def\bc{\begin{center}}
\def\ec{\end{center}}

\def\eg{{\it e.g.}}

\def\to{\rightarrow}

\def\gsim{\mathrel{\mathpalette\atversim>}}

\def\bc{\begin{center}}
\def\ec{\end{center}}

\def\gsim{\mathrel{\rlap{\lower4pt\hbox{\hskip1pt$\sim$}}

    \raise1pt\hbox{$>$}}}       %greater than or approx. symbol

\def\gsim{\mathrel{\rlap{\lower4pt\hbox{\hskip1pt$\sim$}}
    \raise1pt\hbox{$>$}}}       %greater than or approx. symbol

\begin{document}
\makeatletter
\def\fmslash{\@ifnextchar[{\fmsl@sh}{\fmsl@sh[0mu]}}
\def\fmsl@sh[#1]#2{%
  \mathchoice
    {\@fmsl@sh\displaystyle{#1}{#2}}%
    {\@fmsl@sh\textstyle{#1}{#2}}%
    {\@fmsl@sh\scriptstyle{#1}{#2}}%
    {\@fmsl@sh\scriptscriptstyle{#1}{#2}}}
\def\@fmsl@sh#1#2#3{\m@th\ooalign{$\hfil#1\mkern#2/\hfil$\crcr$#1#3$}}
\makeatother
%\baselineskip 24pt

%%%%%%%%%%%%%%%%%%%%%%%%%%%%%%%%%%%%%%%%%%%%%%%%%%%%%%%%%%%%%%%%%
%%%
%%%                      TITLE PAGE
%%%
%%%%%%%%%%%%%%%%%%%%%%%%%%%%%%%%%%%%%%%%%%%%%%%%%%%%%%%%%%%%%%%%%
\thispagestyle{empty}
\begin{titlepage}
\boldmath
\begin{center}
  \Large {\bf A review of Quantum Gravity at the Large Hadron Collider}
    \end{center}
\unboldmath
\vspace{0.2cm}
\begin{center}
{
{\large Xavier Calmet}\footnote{x.calmet@sussex.ac.uk}
}
 \end{center}
\begin{center}
{\sl Physics and Astronomy, 
University of Sussex,  \\ Falmer, Brighton, BN1 9QH, UK 
}
\end{center}
\vspace{\fill}
\begin{abstract}
\noindent

The aim of this article is to review the recent developments in the phenomenology of quantum gravity at the Large Hadron Collider. We shall pay special attention to four-dimensional models which are able to lower the reduced Planck mass to the TeV region and compare them to models with a large extra-dimensional volume. We then turn our attention to reviewing the emission of gravitons (massless or massive) at the LHC and to the production of small quantum black holes.

\end{abstract}  
\end{titlepage}

%\pacs{}

%%%%%%%%%%%%%%%%%%%%%%%%%%%%%%%%%%%%%%%%%%%%%%%%%%%%%%%%%%%%%%%%
%%%
%%%                     INTRODUCTION
%%%
%%%%%%%%%%%%%%%%%%%%%%%%%%%%%%%%%%%%%%%%%%%%%%%%%%%%%%%%%%%%%%%%

\newpage

\section{Introduction}	
One of the main challenges of modern theoretical physics is the unification of quantum mechanics and general relativity. This is a particularly difficult task given the lack of experimental guidance. Indeed, one traditionally expects that quantum gravitational effects will become relevant at an energy scale corresponding to the reduced Planck mass $\bar M_P$ or some $2.43 \times 10^{18} \ {\rm GeV}$. However it has been realized  \cite{ArkaniHamed:1998rs,Antoniadis:1998ig,Randall:1999ee} some 10 years ago, that if there are  large extra-dimensions, i.e., a large extra-dimensional volume, the scale at which gravity becomes strong, $\mu_\star$, could be much smaller than naively expected. It was recently realized that even in four-dimensions, $\mu_\star$ could be much smaller than $\bar M_P$ if there is a large hidden sector of particles that only interact gravitationally with the standard model  \cite{Calmet:2008tn,Dvali}. The aim of this article is to review these recent developments. The large extra-dimensional scenarios have been covered extensively in different excellent reviews, see e.g. \cite{Giudice:2008zza,Morrissey:2009tf}, and we shall thus focus mainly on the new four-dimensional models. We shall however briefly review the large extra-dimensions scenarios in order to compare them with the new four-dimensional models.

General Relativity is remarkably successful on macroscopic scales and it describes all observations and experiments performed on distances from cosmological scales to distances of 10 $\mu$m, see e.g. \cite{review} for a review. More experiments are planned to probe General Relativity on yet shorter scales studying deviations of Newton's potential while, as we shall see, the Large Hadron Collider will probe gravity in the TeV region. Astrophysics also allows us to probe short distance modifications of general relativity and we shall briefly review these bounds.

The main theoretical motivation for lowering the reduced Planck mass to the TeV region is that it would explain the energy gap between the electroweak scale and the scale of quantum gravity. In these frameworks, all interactions of nature become comparable in strength at a few TeV. It should be noted, however, that models with a large extra-dimensional volume do not solve the notorious hierarchy problem, but merely reformulate it in terms of geometry. On the other hand, models with a large hidden sector do solve the hierarchy problem.

In the first part of this review, we shall describe the different theoretical frameworks which can lead to strong gravitational effects in the TeV region. Starting first with four-dimensional models. We shall then briefly review models with a large extra dimensional volume. Then, we shall review the bounds on the different scenarios and study their phenomenology at the Large Hadron Collider.  The signatures which have been emphasized in the literature are the production of massless or massive, i.e. Kaluza-Klein, gravitons or the production of small black holes. We shall briefly give an overview of the state of the art of small black hole production at colliders. 

\section{Models for low scale quantum gravity}
In this section we shall review in detail four-dimensional models which can bring the energy scale at which gravity becomes strong to the TeV region. We shall then briefly review models with a large extra-dimensional volume to compare them to the four-dimensional ones.
\subsection{Models in four dimensions}
\subsubsection{Renormalization of Newton's constant}

Let us consider matter fields of spin 0, 1/2 and 1 coupled to gravity:
\begin{eqnarray} \label{effac1}
S[g,\phi, \psi, A_\mu]&= -\int d^4x \sqrt{-\det(g)} & \left   (\frac{1}{16\pi G_N} R+  \frac{1}{2} g^{\mu\nu} \partial_\mu \phi 
 \partial_\nu \phi + \xi R \phi^2 + \right .  \\ && \nonumber \left . +  e \bar \psi i \gamma^\mu D_\mu \psi + \frac{1}{4} F_{\mu\nu}F^{\mu\nu} \right)
\end{eqnarray}
where $e$ is the tetrad, $D_\mu=\partial_\mu + w^{ab}_\mu \sigma_{ab}/2$ and $ w^{ab}_\mu$ is the spin connection which can be expressed in terms of the tetrad, finally $\xi$ is the non-minimal coupling.

\begin{figure}
%[htp]
\centering
\includegraphics[ width=3in]{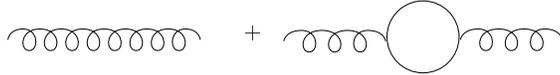}
\caption{Contributions to the running of Newton's constant}\label{Figure2CHS}
\end{figure}

 Let us first consider the contribution of the real scalar field with a $\xi=0$ to the renormalization of the Planck mass.  
Consider the gravitational potential between two heavy, non-relativistic sources, which arises through graviton exchange (Figure \ref{Figure2CHS}). 
The leading term in the gravitational Lagrangian is 
$G^{-1}_N R \sim G^{-1}_N h \Box h$ with $g_{\mu \nu} = \eta_{\mu \nu} + h_{\mu \nu}$. By not absorbing $G_N$ into the definition of the small fluctuations $h$ we can interpret quantum corrections to the graviton propagator from the loop in Figure \ref{Figure2CHS} as a renormalization of $G_N$. Neglecting the index structure, the graviton propagator with one-loop correction is
\begin{equation}
\label{oneloopcorrection}
D_h (q) ~\sim~ \frac{i\,G_N}{q^2} ~+~ \frac{i\,G_N}{q^2} \Sigma \frac{i\,G_N}{q^2} ~+~ \cdots ~ ,
\end{equation}
where $q$ is the momentum carried by the graviton. The term in $\Sigma$ proportional to $q^2$ can be interpreted as a renormalization of $G_N$, and is easily estimated from the Feynman diagram:
\begin{equation}
\label{loop}
\Sigma ~\sim~ -i q^2 \int^\Lambda d^4p~ D(p)^2 p^2 ~+~ \cdots ~ ,
\end{equation}
where $D(p)$ is the propagator of the particle in the loop. In the case of a scalar field the loop integral is quadratically divergent, and by absorbing this piece into a redefinition of G in the usual way one obtains an equation of the form
\begin{equation}
\label{RG1}
\frac{1}{G_{N, \rm ren}} ~=~ \frac{1}{G_{N, \rm bare}} + c \Lambda^2~,
\end{equation}
where $\Lambda$ is the ultraviolet cutoff of the loop and $c \sim 1/16 \pi^2$. $G_{N, \rm ren}$ is the renormalized Newton constant measured in low energy experiments. This result can be derived rigorously using the heat kernel method (see appendix B).

 The running of the reduced Planck mass due to  non-minimally coupled real scalar fields, Weyl fermions and vector bosons can be deduced from the running of Newton's constant  \cite{Calmet:2008tn,Atkins2} see also \cite{Larsen:1995ax,Kabat:1995eq,Vassilevich:1994cz}:
\begin{eqnarray}
\bar M(\mu)^2=\bar M(0)^2-\frac{1}{16 \pi^2} \left  (\frac{1}{6} N_l + 2 \xi N_\xi \right ) \mu^2
\end{eqnarray}
where $\mu$ is the renormalization scale and $N_l=N_S+N_F-4 N_V$ where $N_S$, $N_F$ and $N_V$ are respectively the numbers of real, minimally coupled, scalar fields, Weyl fermions and vector bosons in the model and $N_\xi$ is the number of real scalar fields in the model with a non-minimal coupling to gravity. Note that the conformal value of $\xi $ in our convention is $1/12$.  The renormalization group equation at one loop for the reduced Planck mass is obtained using the heat kernel method  which preserves the symmetries of the problem.

The scale at which quantum gravitational effects become strong,  $\mu_\star$,  follows from the requirement that the reduced Planck mass at this scale $\mu_\star$ be comparable to the inverse of the size of the fluctuations of the geometry, in other words, $\bar M(\mu_\star) \sim \mu_\star$.  One finds \cite{Atkins:2010eq}:
\begin{eqnarray}
\mu_\star=\frac{\bar M(0)}{\sqrt{1+\frac{1}{16 \pi^2} \left  (\frac{1}{6} N_l + 2 \xi N_\xi \right )}}.
\end{eqnarray}
Clearly the energy scale at which quantum gravitational effects become relevant depends on the number of fields present in the theory and on the non-minimal coupling $\xi$. While minimally coupled spin 0 and spin $1/2$ fields lower $\mu_\star$, spin 1 fields increase the effective reduced Planck mass and non-minimally coupled scalar fields can increase or lower $\mu_\star$ depending on the algebraic sign of $\xi$. The contribution of the graviton is a $1/N$ effect and very small if $N$ is reasonably large. 

There are different ways to obtain $\mu_\star=1$ TeV. The first one is to introduce a large hidden sector of scalars and/or Weyl fermions with some $10^{33}$ particles. The other one is to consider a real scalar field that is non-minimally coupled with a $\xi\sim 10^{32}$. As we shall both choice lead to a violation of unitarity at an energy scale below $\mu_\star$.

\subsubsection{Black Hole argument}
It has been proposed in  \cite{Dvali}  that a large hidden sector leads to a gravitational cutoff below the naive Planck mass in four dimensions. They consider a model with $N$ species and imposes an exact  discrete $Z_2^{N}=Z_2^{1} \times Z_2^{2} \times ... \times Z_2^{(N)}$ symmetry under the independent sign flips of the fields $\phi_j \to -\phi_j$. They then consider the minimal size black hole carrying the maximum possible discrete charge. Its mass is given by $M_{BH} = N \Lambda$, $\Lambda$ being the mass of a particle charged under  $Z_2^{(N)}$. They argue that the information about the $Z_2^N$  charge carried by the black hole must be conserved in the decay of the black hole. For a black hole with a  Hawking temperature $T_H$, the probability of the emission of a heavy particle of mass  $\Lambda$ greater than $T_H$ is exponentially suppressed by a Boltzmann factor  $\exp(- \Lambda/T_H)$. Thus, the black hole with $N$ units of the $Z_2^N$ charge, can start
emitting $N$ species particles, only after its temperature has dropped to $T_H \sim \Lambda$. At this point, the
mass of the black hole is $M^\star_{BH}\sim M_P^2/\Lambda$. Starting from this moment, the black hole can start emitting some of the $N$ species of particles. However, by conservation of energy, the maximum number of particles that can be emitted by the black hole is
$n_{\rm max}\sim M_P^2/\Lambda^2$. These	states should carry the same $Z_2^N$ -charge as the original	$N$ particles. Hence, $n_{\rm max} =N$. The only way to avoid an inconsistency is that the squared Planck mass scales as $N$. This summarizes the argument presented in \cite{Dvali}.

This argument suffers from a number of difficulties. First of all, quantum gravity does not necessarily preserve discrete symmetries \cite{Banks:1989zw,Holman:1992us} and the black hole considered above could evaporate via channels that violate the conservation of $Z_2^N$ symmetries. Secondly, it relies on a spin independent entropy for the black hole. This choice is not unique. It is likely to be the reason for the discrepancy between the running of the Planck mass argument and the black hole argument for spin 1 particles.  Indeed the black hole argument is not able to differentiate between the different spins as the renormalization group argument does. Thirdly, the running of the Planck mass argument clearly shows that the discrete symmetry is not necessary. As one shall see shortly, this model also suffers from issues related to the unitarity of the S-matrix. It should be emphasized that the model based on a renormalization of the Planck mass could be seen as a valid realization of the idea proposed in \cite{Dvali}.

\subsubsection{Unitarity issues}

The action (\ref{effac1}) can be linearized using  $g_{\mu\nu}=\eta_{\mu\nu}+\sqrt{2}h_{\mu\nu}/\bar M_P + {\cal O}(\bar M_P^{-2})$, where the scale, i.e the reduced Planck mass, appearing in this expansion is fixed by the requirement that the kinetic term of the graviton be  canonically normalized. One obtains the following Lagrangian
\begin{eqnarray} \label{efflag1}
L&=& -\frac{1}{4} h^{\mu\nu} \square h_{\mu\nu} +\frac{1}{4} h \square h -\frac{1}{2} h^{\mu\nu} \partial_\mu \partial_\nu h +   \frac{1}{2}h^{\mu\nu} \partial_\mu \partial_\alpha h_\nu^\alpha \\ \nonumber &&  - \frac{\sqrt{2}}{\bar M_P} h^{\mu\nu}  T_{\mu\nu} + {\cal O}( \bar M^{-2}_P)
\end{eqnarray}
where $T^{\mu\nu}$ is the energy-momentum tensor corresponding to the matter content of the theory. This action can be regarded as an effective action valid up to $\bar M_P\sim 2.43 \times 10^{18} \ \mbox{GeV}$.  Traditionally one expects that gravitational interactions become strong above this energy scale and  the metric should not be linearizable at higher energies. In that sense one can consider that linearized General Relativity is an effective theory valid up to an energy scale corresponding to the reduced Planck mass.

In \cite{Atkins:2010eq}  a criteria was introduced  for the validity of linearized General Relativity coupled to matter. The scale at which unitarity is violated, $E_\star$, in the gravitational scattering of particles of spin 0, 1/2 and 1 needs to be compared to the scale at which quantum gravitational effects become strong ,i.e. $\mu_\star$. The tree level amplitudes had been obtained previously in \cite{Han:2004wt}. Using this criteria, one derives a bound on the particle content of a model coupled to linearized General Relativity. One finds  \cite{Atkins:2010eq}:
\begin{eqnarray} \label{bound1}
\frac{N_S}{3}+N_F+4 N_V \le  160 \pi
\end{eqnarray}
using the $J=2$ partial wave and 
\begin{eqnarray} \label{bound2}
(1+12 \xi)^2 N_S \le  96 \pi
\end{eqnarray}
using the $J=0$ partial wave. These bounds are obtained considering gravitational scattering of the type  $ 2 \phi_i \to 2 \phi_j$ with $i\ne j$ which are $s$-channel processes. Imposing different ingoing and outgoing particles insures the absence of $t$ and $u$-channels. These bounds are thus valid for $N_i>1$.

We  can obtain the following bound on the non-minimal coupling of the scalar field to the Ricci scalar 
\begin{eqnarray} \label{boundxi}
-\frac{ 4\sqrt{6 \pi N_S}  +N_S }{12 N_S} \le \xi \le \frac{ 4\sqrt{6 \pi N_S}  -N_S }{12 N_S},
\end{eqnarray}
using the $J=0$ amplitude and requesting that the effective action remains valid up to the reduced Planck mass, i.e. by setting  $\sqrt{s}=\bar M_P$. Note that this bound is valid for $N_S>1$. For $N_S=1$ there is a cancellation of the terms proportional to $\xi$ that grow with energy  \cite{Atkins2}.

Clearly in models with a large hidden sector or with a large non-minimal coupling, unitarity is violated below the reduced Planck mass and some new physics needs to be introduced to restore unitarity  \cite{Atkins:2010eq}. One option would be to embed these models in string theory models with a string scale below the Planck mass in the hope that non local stringy effects restore unitarity up to the reduced Planck mass $\mu_\star$. This has important experimental consequences. The first signal of these models at a collider is unlikely to be of gravitational nature, but could rather be a sign of the non-local (i.e. extension in space) nature of leptons and quarks.

\subsection{Large extra-dimensions}

These models have been extensively studied over the last ten years.  Models with large extra dimensions  assume that standard model excitations are confined to a 3+1 sub-geometry, and employ the following trick. The higher dimensional action is of the form 
\begin{eqnarray} \label{acextradim}
S=\int d^4x \, d^{d-4}x' \, \sqrt{-g}  \left( M_\star^{d-2} \, {\cal R} + \cdots   \right) 
\end{eqnarray}
and the effective 3+1 gravitational energy scale (Planck scale) is given by
 \begin{eqnarray} 
M_p^2 = M_\star^{d-2} V_{d-4}
\end{eqnarray}
where $V_{d-4}$ is the volume of the extra dimensions. By taking $V_{d-4}$ large, $M_p$ can be made of order $10^{19}$ GeV while $M_\star \sim$ TeV, at the cost of some strong dynamical assumptions about the geometry of space-time.
 There are different realizations of this idea. In the ADD, which stands for Arkani-Hamed, Dimopoulos and Dvali, \cite{ArkaniHamed:1998rs,Antoniadis:1998ig} brane world model,  the particles of the standard model are assumed to be confined to a three dimensional surface, called  a brane, whereas gravity can propagate everywhere i.e. on the brane and in the extra-dimensional volume called the bulk. The number of extra-dimensions is not determined from first principles. In the version proposed by Randall and Sundrum (RS) \cite{Randall:1999ee}, a five-dimensional space-time is considered with two branes. In the simplest version of the RS model, the standard model particles are confined to the so-called IR brane while gravity propagates in the bulk as well. One of the main difficulties of models with large extra-dimensions is that of proton decay. 
 In the case of RS, it was later on proposed to allow the leptons and quarks to propagate in the bulk to suppress proton decay operators  \cite{Huber:2003tu}.

\subsubsection{Unitarity issues}
Models with large extra-dimensions also typically suffer from unitarity problems, see  e.g. \cite{Grzadkowski:2007zz,Kachelriess:2000cb,DeCurtis:2003zt,Pelaez:2005hd} for some examples. These examples can be generalized. It was shown in \cite{Atkins2} that in models with a large extra-dimensional volume, unitarity is violated below the scale at which gravity becomes strong in the scattering of particles of the standard model via Kaluza-Klein excitations of the graviton. The problem appears because of the large number of Kaluza-Klein excitations of the graviton.  
The calculation of the unitarity bound for these models   is  very similar to that described previously for the unitarity bound relevant to the large hidden sector model.
In large extra-dimensional models,  the fundamental Planck scale is in the TeV region and because the volume is large there are approximately $N_{KK}=10^{32}$ Kaluza Klein (KK) gravitons with masses below 1 TeV. Scattering between the matter content of the model can now take place via exchange of any one of this very large number of KK modes and it is found  \cite{Atkins2}  that the $J=2$ partial wave, in the massless limit, is unaltered to that found for massless gravitons. One finds  that the $J=2$ partial wave amplitude is given by \cite{Atkins2}:
\begin{eqnarray} 
\left |a_2 \right |= \frac{1}{320 \pi} \frac{s}{\bar M_P^2}N_{KK} N.
\end{eqnarray}
For the case of the standard model coupled to $10^{32}$ KK gravitons we find that at $\sqrt{s}=1$ TeV, $|a_2| \sim 1.6$ and unitarity is violated $E^\star_{\rm CM}= 561$ GeV which is clearly below the scale at which gravity is supposed to become strongly coupled.

 As in the large hidden sector four dimensional models, these models could be embedded in string theoretical models such as little string theory at a TeV \cite{Antoniadis:2001sw,Calmet:2007je}. The same experimental consequences as for the four dimensional models follow.

\section{Phenomenology}
The main signatures of models with a low scale reduced Planck mass are the productions at colliders of gravitons which are massless in the case of a large hidden sector \cite{Calmet:2009gn,Calmet:2009yw} or massive  \cite{Han:1998sg,Mirabelli:1998rt,Giudice:1998ck} in the case of  large extra-dimensional models,  and of small black holes  \cite{Dimopoulos:2001hw,Giddings:2001bu,Feng:2001ib,Anchordoqui:2003ug,Anchordoqui:2001cg,Anchordoqui:2003jr,Meade:2007sz,Calmet:2008dg,Kanti:2004nr} in both cases. 

\subsection{Bounds on the Planck mass}
\subsubsection{Four dimensions}
The most serious bound \cite{Calmet:2008rv} to date on the four-dimensional reduced Planck mass comes from the cosmic ray experiment Akeno Giant Air Shower Array (AGASA) \cite{Yoshida:2001pw,Inoue:1999cn}.  Anchordoqui et al. \cite{Anchordoqui:2001cg} had proposed to use Earth skimming neutrinos \cite{Feng:2001ue} to probe new strong interactions in the neutrino sector. The idea is that within the standard model there is a small chance for a neutrino that interacts with a nuclei in the Earth crust to form a tau-lepton which can escape the crust and create a shower in the atmosphere. AGASA is expected to see about three of these events per year. A deficit in these events would signal a new strong interaction in the neutrino sector. In particular if the scale of quantum gravity is in the TeV region, small black holes could be formed in the Earth crust. Gravity being democratic, these small black holes would decay uniformly to all allowed final states and there would thus be a deficit of tau-leptons and hence of showers. The theory of small black hole formation is reviewed below. AGASA data leads to a bound on the four dimensional Planck mass of the order of 488 GeV. It should be stressed that there are sizable uncertainties in the derivation of this bound. It depends on the nature of the most energetic cosmic rays and on their flux. Limits from the Tevatron have not be studied in detail, but are not expected to provide a significantly tighter bound. On the other hand the LHC is expected to set a limit of up to 5 TeV on the reduced Planck mass with a luminosity of 100 fb$^{-1}$ and  a center of mass energy of 14 TeV \cite{Calmet:2009gn}.

\subsubsection{Large extra-dimensions}

There are different sources of limits on the number of extra-dimensions and the scale of quantum gravity for models with a large extra-dimensional volume. In the case of RS, there is a mass gap between the graviton and its first Kaluza-Klein excitation which typically has a mass of the order of 1 TeV, the bound on the scale of quantum gravity is of the order of 1 TeV if the fermions, the gauge bosons and the Higgs boson are confined to the IR brane. If they are allowed to propagate in the bulk, the limit becomes much tighter and typically of the order of 10 TeV \cite{Huber:2003tu}.

In the case of ADD, the LEP and the Tevatron set limits of the order of 1 TeV \cite{LEP,Landsberg:2004mj} (for $n$=1 to 7 extra-dimensions) from the emission of Kaluza-Klein modes of the graviton and Drell-Yan processes. Astrophysical measurements, in particular limits on supernovae cooling, neutron stars reheating and the absence of a diffuse $\gamma$-ray background, allow one to set limits on the reduced Planck mass in models\cite{Hannestad:2001jv}. 
The bounds on the reduced Planck mass are of the order of $10^3$ TeV for $n=2$, $10^2$ TeV for $n=3$ and 5 TeV for $n=4$ where $n$ stands for the number of extra-dimensions. The case of $n=1$ and $\mu_\star= 1$ TeV is ruled out by solar system physics. $n=2$ and $\mu_\star=1$ TeV leads to modifications of Newton's $1/r$ potential on distances of 0.2 mm and is now ruled out. AGASA (see above) typically leads to bounds in the few TeV region for $n$=1 to 7 extra-dimensions and is the only source of bounds for $n\ge 5$. However the same caveat as that mentioned above for $n=0$ applies. Typically, only $n\ge4$ ADD models are  relevant for LHC physics.

\subsection{Quantum gravity at the Large Hadron Collider}
\subsubsection{Solving the unitarity problem}
Models with a large hidden sector or large extra-dimensional volume are designed to lower the Planck mass in the TeV region. As mentioned above, they however typically have problems with unitarity below the scale $\mu_\star$. One solution could be to embed these models in a non-local theory of gravity such as string theory with a string scale below the Planck mass. A specific scenario could be little string theory at a few TeV \cite{Antoniadis:2001sw,Calmet:2007je}. Although there is no real compelling solution to the unitarity problem at this point, one can however deduce that if the reduced Planck mass is in the TeV region, the first signals the LHC will find are not of gravitational nature, but rather linked with the physics which cures the unitarity problem of these models. In the case of little string theory, the firsts signals would reveal the stringy nature of the particles \cite{Anchordoqui:2007da,Anchordoqui:2003jr}.

The lowest $\mu_\star$ one can archive without having a violation of unitarity below that scale is 14 TeV \cite{Atkins2}. This can be obtained in a four-dimensional model with a single real scalar field strongly non-minimally coupled to the Ricci scalar. The renormalization of the reduced Planck mass leads to a $\mu_\star=14$ TeV for $\xi=2.3 \times 10^{30}$. 

\subsubsection{Graviton emission at the LHC}
At the LHC, the production of jets with large $E_T$ recoiling against a graviton $G$ can 
arise from the  parton subprocesses  $q+ \bar q \to G + g$, $q+ g \to q+ G$,
 $\bar q + g \to \bar q + G$, and $g+g \to g+ G$. Using the Feynman rules given in appendix A for linearized four-dimensional general relativity coupled to the standard model, the leading order contributions at the parton level have been calculated in \cite{Calmet:2009gn}.  All quarks are treated as being massless.
 The polarization and color averaged cross section for $q + \bar q \to g +G$ is given by
 \begin{eqnarray}
\frac{d \sigma}{d \cos \theta} = \frac{g_s^2}{ 72 \pi \bar M(\mu)^2}
\end{eqnarray}
where $g_s$ is the strong coupling constant, $\bar M(\mu)$ is the reduced Planck Mass and where $s$ and $t$ are the Mandelstam variables: $t= -1/2 s (1 - \cos \theta)$.
The  cross sections for $q + g \to q + G$ and $\bar q + g \to \bar q + G$  are given by
 \begin{eqnarray}
\frac{d \sigma}{d \cos \theta} = - \frac{g_s^2 t}{ 192 \pi \bar M(\mu)^2 s}.
\end{eqnarray}
The corresponding matrix element can be obtained using crossing symmetry from that of the transition $q +\bar q \to G + g$.
Finally the cross section for  $g+ g \to g + G$ is given by:
 \begin{eqnarray}
\frac{d \sigma}{d \cos \theta} = - \frac{3 g_s^2 (s^2+s t + t^2)^2}{ 64 \pi \bar M(\mu)^2 s^3 t (s+t)}.
\end{eqnarray}
The total cross sections  are given by:
 \begin{eqnarray}
 \sigma(q\bar q \to gG) &=&  \frac{g_s^2}{ 36 \pi \bar M(\mu)^2}, \\
 \sigma(q g  \to q G)&=&  \frac{g_s^2 }{ 192 \pi \bar M(\mu)^2},\\
 \sigma(\bar q g  \to \bar q G)&=&\sigma(q g  \to q G), \\
  \sigma(gg \to g G)&=&\frac{17 g_s^2 }{ 128 \pi \bar M(\mu)^2}.
\end{eqnarray}
In what follows the energy scale $\mu$ is identified with the partonic center of mass energy $\sqrt{\hat s}$. For collisions with $\sqrt{\hat s} < \mu_* \sim 1$ TeV, quantum gravity contributions are so weak ($\bar M(\mu) \sim 10^{18}$  GeV for $\sqrt{\hat s} < \mu_*$) that the cross sections go to zero fast. However for $\sqrt{\hat s} > \mu_*$, $\bar M(\mu)\sim$1 TeV and gravitons will be produced. 
The running of the Planck mass can be implemented with a Heaviside step function in the cross section, i.e. the Planck mass for collisions at the parton level with $\sqrt{\hat s}>1$ TeV is given by $\bar M(\mu_*)=\bar \mu_*= 1$ TeV,  but for less energetic parton level collisions $\bar M \to \infty$ is assumed. Since most of the running takes place close to $\mu_\star$, this is a very accurate approximation.
Because the parton level cross sections are independent of the center of mass energy, one can write
 \begin{eqnarray}
\sigma(PP\to Graviton + jets)&=& 
\int^1_{\bar \mu_*^2/s} du \int^1_u \frac{dv}{u}
 \\ \nonumber &&
 \biggl (   \sigma(q \bar q \to gG) 
 \sum_{\begin{subarray}{l}  i=1..6 \\  j=-1..-6 \end{subarray}}  f_i(v,Q) f_j(u/v,Q) \\ \nonumber && + \sigma(q g  \to q G)
  \sum_{i=1..6} f_i(v,Q) f_0(u/v,Q)   \\ \nonumber &&
+  \sigma(\bar q g  \to \bar q G)
 \sum_{i=-1..-6} f_i(v,Q) f_0(u/v,Q)  \\ \nonumber && 
 +  \sigma(gg \to g G)
  f_0(v,Q) f_0(u/v,Q)  \biggl).
   \end{eqnarray}
The cross section for proton+proton $\to$ Graviton + jets at a center of mass of 14 TeV is  approximatively $4.3 \times 10^4$ fb \cite{Calmet:2009gn}. Obviously the graviton is not detectable and appears as missing energy and  the signature for the emission of  a graviton is then proton + proton $\to$ jets + missing energy.
\subsubsection{KK Graviton emission at the LHC}
 The corresponding cross-sections have been obtained by different groups  \cite{Mirabelli:1998rt}. One finds \begin{eqnarray}
\frac{d\sigma}{d \cos\theta}(q + \bar q \to g + G_{KK})  &=& \frac{1}{144 \pi}
 \frac{g_s^2 }{\bar M^2}\frac{1}{1-m^2/s}\Biggl[(2 - \frac{4ut}{(s-m^2)^2}) 
  \left(1 + \bigl(\frac{m^2}{s}\bigr)^4\right) \\ \nonumber
     &&  + \left(2 \frac{ (s-m^2)^2}{4 ut} - 5 
      + 4 \frac{4ut}{(s-m^2)^2}\right)\frac{m^2}{s } \left(1 +  
   \bigl(\frac{m^2}{s}\bigr)^2\right) + 
   \\ \nonumber
     && + 6 \left( \frac{u-t}{s-m^2}\right)^2
                          \bigl(\frac{m^2}{s}\bigr)^2
                                          \Biggr] \ ,
\end{eqnarray}
where $s,t,u$ are the Mandelstam variables with the usual definitions: $t,u = -1/2 s (1-m^2/s)(1\mp 
\cos \theta)$ for the cross section $q + \bar q \to g + G_{KK}$ where $G_{KK}$ is a Kaluza-Klein graviton.
The cross section for $q+g \to q+ G_{KK}$ 
 can be obtained from this expression 
using the crossing symmetry $s \leftrightarrow t$:
\begin{eqnarray}
\frac{d\sigma}{d \cos\theta}(q+g \to q+ G_{KK})  &=&  
\frac {g_s^2}{384 \pi \bar M^2} \frac{(-t/s)(1-m^2/s)}{(1-m^2/t)^2} \times \\
\nonumber
     && 
 \times \Biggl[(2 - \frac{4us}{ (t-m^2)^2}) 
  \left(1 + \bigl(\frac{m^2}{t}\bigr)^4\right) \\ \nonumber
     && + \left(2 \frac{ (t-m^2)^2}{4 us} - 5 
      + 4 \frac{4us}{(t-m^2)^2}\right)\frac{m^2}{t }\left(1 +  
   \bigl(\frac{m^2}{t}\bigr)^2\right) +\\ \nonumber
     && + 6 \left( \frac{s-u}{t-m^2}\right)^2
                          \bigl(\frac{m^2}{t}\bigr)^2
                                          \Biggr] \ .
\end{eqnarray}
As in the massless case, the cross section for $\bar q +g \to \bar q + G_{KK}$ is also
 the same as that of $q +g \to q + G_{KK}$. 
  For the process
 $g+g \to g+ G_{KK}$, one finds 
 \begin{eqnarray} \nonumber
\frac{d\sigma}{d \cos\theta}(g+g \to g+ G_{KK})&=& \frac{3}{16}
      \frac{\pi \alpha_s G_N}{(1-m^2/s)(1-\cos^2\theta)}\Biggl[(3 + 
   \cos^2\theta )^2\left(1 + \bigl(\frac{m^2}{s}\bigr)^4\right) \\ 
     && - 4 ( 7 + \cos^4\theta)\frac{m^2}{s }\left(1 +  
           \bigl(\frac{m^2}{s}\bigr)^2\right) 
            \\ \nonumber
     && 
          + 6 (9 - 2  \cos^2\theta +  \cos^4\theta)  
                      \bigl(\frac{m^2}{s}\bigr)^2  
                                          \Biggr] \ .
\end{eqnarray}
In the massless limit $m\to 0$, the cross-sections for the production of massive Kaluza-Klein modes match the cross-sections of the massless graviton described above.

\subsubsection{ Massless versus Kaluza-Klein gravitons at the LHC}
\begin{figure}
\centering
\includegraphics[width=8cm,angle=90]{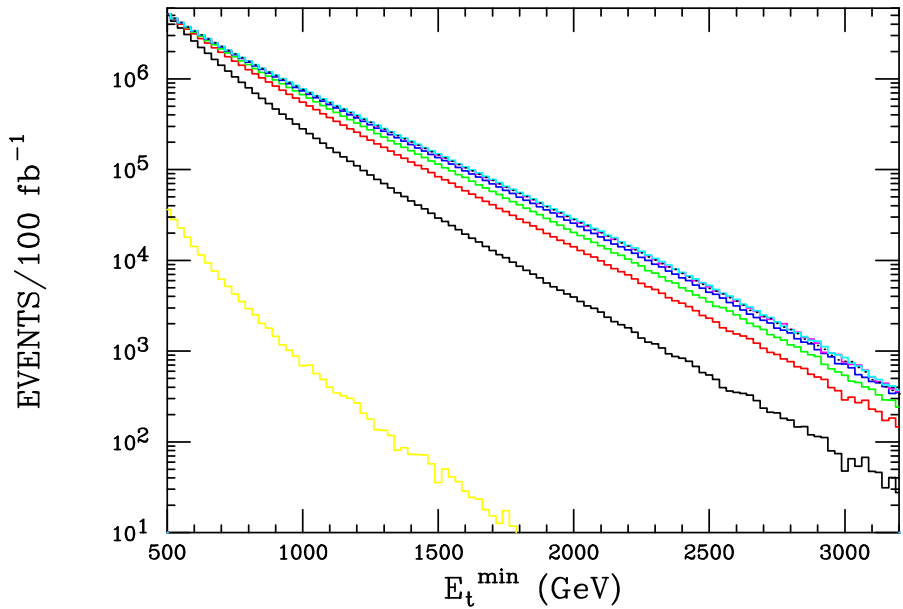}
\caption{This figure shows the shape of the $E_T^{min}$ distribution counting the number of monojet events at the 14 TeV LHC assuming a luminosity 
of 100 fb$^{-1}$ 
above a $E_T^{min}$ cut of 500 GeV and requiring a central jet $|\eta_j| <3$. The yellow histogram is the expected standard model 
background as discussed in the text while the red and higher histograms are for the ADD model with the number of extra dimensions being 2,3,4, etc. 
The lower solid black histogram is for the four dimensional model with $M_P$=1 TeV. The ADD results were in each case adjusted by varying their 
associated Planck scale to produce the same result as does the four dimensional model at $E_T^{min}=500$ GeV in order to show the relative shapes for 
these distributions.}
\label{Figure1TR}
\end{figure}

Using the partonic cross sections above, a modified version of the code developed for \cite{Rizzo:2008vr} has been employed \cite{Calmet:2009yw} to generate events  for the 14 TeV LHC for both the ADD and the four dimensional models as well as for the 
the standard model background. Here, following Ref.{\cite {hinch}}, one expects this background 
to be dominated by the production of $Z$ plus a single jet with the $Z$ decaying into pairs of neutrinos. 
This background can be much reduced by requiring a missing energy and/or jet energy cut of at least 500 GeV 
and demanding that the jet be central $|\eta_j|<3$. The results of our direct comparison of the ADD predictions with those of the four 
dimensional model assuming, \eg, that $\bar M_P=1$ TeV, can be found in Fig. \ref{Figure1TR}. Here one sees that the two 
new physics models predict monojet $E_T$ distributions which are reasonably dissimilar in overall shape. The falling 
four dimensional model monojet spectrum above the $\sqrt {\hat s}=\bar M_P$ threshold is seen to be somewhat softer 
than the corresponding ADD model prediction for any number of extra dimensions. 

\begin{figure}[htp]
\centering
\includegraphics[width=8cm,angle=90]{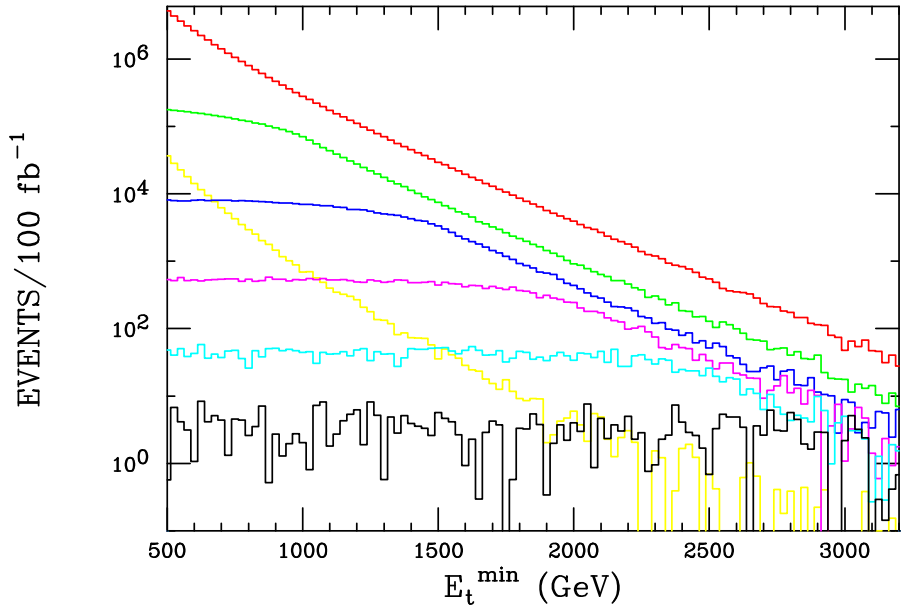}
\caption{This figure for the 14 TeV LHC shows the event rate for the standard model jet+missing energy background as a function of the cut on the jet $E_T$ 
in yellow as well as the four dimensional model predictions for the cases $M_{P}$=1(2,3,4,5,6) TeV from top to bottom in red, green, blue,...  
and requiring that the existence of a threshold at $\sqrt {\hat s}= M_P$. From this figure, one can deduce that 
the search reach for the four dimensional model at the LHC is $\simeq$ 5 TeV for a luminosity =100 fb$^{-1}$. The 
shape of the signal histograms with the requirement above are quite different than those for ADD due to the vanishing of the cross 
section at small $\hat s$. Note the shape change at $E_T^{min}=0.5 M_P$ which is a result of this cross section threshold that is  
absent in the ADD model.}
\label{Figure2TR}
\end{figure}

\begin{figure}
\centering
\includegraphics[width=8cm,angle=90]{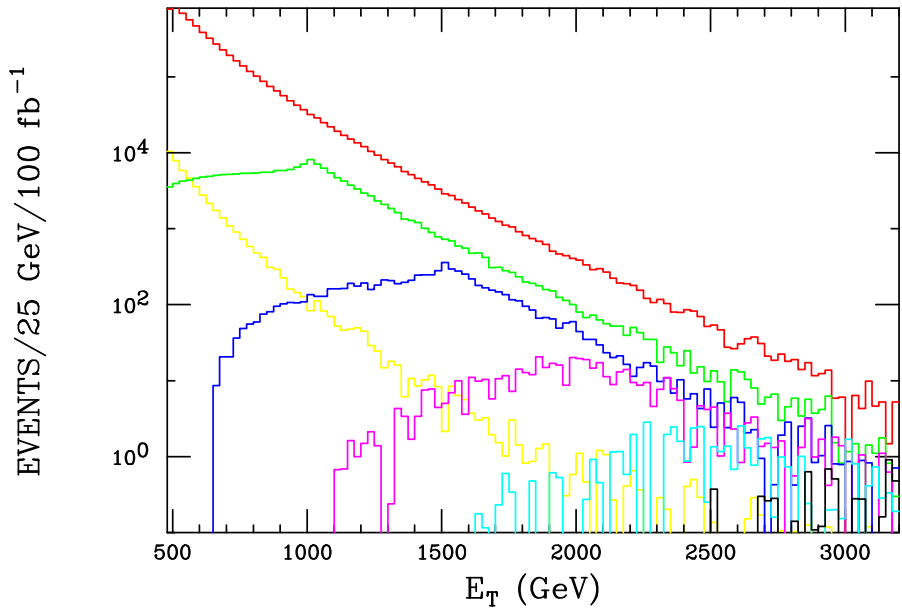}
\caption{This figure shows the monojet $E_T$ distributions at the 14 TeV the LHC assuming a luminosity of 100 fb$^{-1}$. The histograms are 
as labeled as in the previous figure. Note the kink-like structure in the four dimensional model distributions at $E_T= 0.5 M_P$ which is a result of 
the cross section threshold that is absent in the ADD model.}
\label{Figure3TR}
\end{figure}

From Figures \ref{Figure2TR} and \ref{Figure3TR} one can obtain an estimate of the search reach for the four dimensional model at the 
14 TeV LHC of $\simeq 5$ TeV assuming an integrated luminosity of 100 $fb^{-1}$. Here one also sees some unusual features in these 
distributions associated with the four dimensional model which are absent from the case of ADD which are due to the cross section 
threshold at $\sqrt {\hat s}=M_P$. Unlike for the ADD case, whose $E_T$ and $E_T^{cut}$ distributions fall monotonically, the 
threshold in the four dimensional model cross section naturally leads to structure in these corresponding distributions. This 
added kink-like structure is a clear aid in distinguishing the predictions of these two classes of models at the LHC as can easily be seen 
from these figures. For the case of the $E_T^{cut}$ distribution one sees that it is essentially flat until the value $E_T^{cut}=0.5M_P$ 
is reached and then falls monotonically. On the other hand, the $E_T$ distribution rises below the value of $E_T=0.5M_P$ at which point a peak occurs. 
At higher values of $E_T$ the distribution fall monotonically. These are quite distinctive indications of a cutoff in the cross section 
at a fixed value of $\sqrt {\hat s}$. This value could be extracted directly from the LHC experimental data.  

\subsubsection{Production of small black holes}
If the scale of quantum gravity is truly as low as a few TeV, the most striking feature of these models is the prediction that colliders such as the LHC may be able to create small black holes \cite{Dimopoulos:2001hw,Giddings:2001bu,Feng:2001ib,Anchordoqui:2003ug,Anchordoqui:2001cg,Anchordoqui:2003jr,Meade:2007sz,Calmet:2008dg}.
\subsubsection{Theory}
In the early days of black hole formation at colliders, the hoop conjecture \cite{hoop} due to Kip Thorne was used as a criteria for gravitational collapse in the collision of two particles head to head. The hoop conjecture states that if an amount of energy $E$ is confined to a spherical region of space-time with a radius $R$ with $R < E$, then that region will eventually evolve into a black hole.  Natural units were used: $\hbar, c$ and Newton's constant are set to unity. It had been known since the works on Penrose (unpublished) and later on by D'Eath and Payne \cite{D'Eath:1992hb} that a small black hole will be formed in the head to head collision of two particles with zero impact parameter. However, the relevant case for the LHC is that of a non vanishing impact parameter. The resolution of the problem was given by Eardley and Giddings \cite{Eardley:2002re} who were able to construct a closed trapped surface. Yoshino and Nambu \cite{Yoshino:2002tx} then showed, that another important effect needs to be included, namely that some significant fraction of the center-of-mass energy is radiated away before the black hole forms. 

\subsubsection{Semi-classical versus non-thermal black holes}
The construction of Eardley and Giddings  \cite{Eardley:2002re} is valid in the limit where the mass of the small black holes and hence the center of mass energy is much larger than the effective reduced Planck mass. The black hole formed in that limit is a  semi-classical black hole. Semi-classical black holes are thermal objects that are expected to decay via Hawking radiations to many particles, typically of the order of 20, after a spin down phase. This final explosion would lead to  a spectacular signature in a detector. It is however now well understood \cite{Anchordoqui:2003ug,Meade:2007sz,Calmet:2008dg}  that it is very unlikely that semi-classical black holes will be produced at the LHC because the center of mass energy is not high enough. The main reasons are that not all of the energy of the partons is available for black hole formation \cite{Yoshino:2002tx}  and the parton distribution functions (PDFs)  tend to fall off very fast. The ratio between the first semi-classical black hole mass and that of the Planck mass can be estimated. In the case of ADD, it is typically taken to be of the order of 5, while it could easily be 20 for RS \cite{Meade:2007sz}.

It has been proposed in  \cite{Calmet:2008dg} to extrapolate the semi-classical black hole into the quantum regime of quantum gravity and to consider  the production of quantum black holes (QBHs)  at the LHC. QBHs are defined as the quantum analogs of ordinary black holes as their mass and Schwarzschild radius approach the quantum gravity scale. QBHs do not have semi-classical space-times and are not necessarily well-described by the usual Hawking temperature or black hole thermodynamics. In other words they are non thermal. In many respects they are perhaps more analogous to strongly coupled resonances or bound states than to large black holes. QBHs presumably decay only to a few particles, each with Compton wavelength of order the size of the QBH. It seems unlikely that they would decay to a much larger number of longer wavelength modes.

It is  assumed that QBHs are defined by three quantities: their mass, spin and gauge charges. Importantly, QBHs can have a QCD, or color, charge. This is not in contradiction with confinement since the typical length scale of QCD, i.e., a Fermi, is much larger than the size of a QBH, e.g., TeV$^{-1}$. The formation and decay of a QBH takes place over a small space-time region -- from the QCD perspective it is a short distance process, and hadronization occurs only subsequently. 

The central assumptions are as follows.

\begin{itemize}

\item[I)] Processes involving QBHs conserve QCD and U(1) charges since local gauge symmetries are not violated by gravity. Note that no similar assumption is made about global charges.
\item[II)] QBH coupling to long wavelength and highly off-shell perturbative modes is suppressed.
\end{itemize}
Assumption (II) is necessary so that precision measurements (e.g., of the anomalous magnetic moment of the muon \cite{Calmet:1976pu}) or, possibly, proton decay do not force the quantum gravity scale to be much larger than the TeV range. It is not implausible that a nonperturbative QBH state couples only weakly to long distance or highly off-shell modes, but strongly to modes of size and energy similar to that of the hole. This is analogous to results obtained for (B+L) violating processes in the standard model: (B+L) violation is exponentially small in low energy reactions, but of order one for energies above the sphaleron mass.

It is hard to imagine that (I) does not hold. Imagine a large Gaussian 3-sphere surrounding the spatial region where QBH formation and decay occurs. By causality, the total flux through this sphere is constant, implying conservation of charge. A consequence of assumption (I) is that QBHs can be classified according to representations of SU(3)$_c$ and U(1)$_{em}$. A QBH state is labeled as QBH$_c^q$. 

Note that Lorentz invariance is not listed as one of the central assumptions. The results will depend significantly on whether one requires that QBH processes correspond to Lorentz invariant, local effective field theory operators (i.e., constructed from the usual standard model fields). There is no known argument in favor of this which is as robust as the one for conservation of gauge charges. The black hole production and decay take place over a small region of space-time with Planckian volume. Whether or not this process can be matched to a local operator in an effective field theory description at larger length scales seems an open question. If quantum gravity does violate Lorentz invariance in QBH processes, the assumption (II) above is expected to be  sufficient to protect low energy physics (e.g., precision measurements) from contamination by these effects.

Finally, in order to obtain quantitative results one needs to assume that QBH production cross-sections can be extrapolated from the cross-section obtained for semiclassical black holes \cite{Anchordoqui:2003ug} (see also \cite{Dimopoulos:2001hw,Giddings:2001bu} for earlier, less elaborated, cross-sections)
\begin{eqnarray}
\sigma^{pp}(s,x_{min},n,M_D) &=& \int_0^1 2z dz \int_{\frac{(x_{min} M_D)^2}{y(z)^2 s}}^1 du \int_u^1 \frac{dv}{v}  \\ \nonumber && \times F(n) \pi r_s^2(us,n,M_D) \sum_{i,j} f_i(v,Q) f_j(u/v,Q)
\end{eqnarray}
where $z=b/b_{max}$, $x_{min}=M_{BH,min}/M_D$,  $n$ is the number of extra-dimensions, $F(n)$ and $y(z)$ are the factors introduced by Eardley and Giddings and by Yoshino and Nambu \cite{Yoshino:2002tx}  (the numerical values from \cite{Yoshino:2002tx} are used) and 
\begin{eqnarray}
r_s(us,n,M_D)=k(n)M_D^{-1}[\sqrt{us}/M_D]^{1/(1+n)}
\end{eqnarray}
where
\begin{eqnarray}
k(n) =  \left [2^n \sqrt{\pi}^{n-3} \frac{\Gamma(3+n)/2}{2+n} \right ]^{1/(1+n)},
\end{eqnarray}
and $M_D$ is the reduced Planck mass.
$M_{BH,min}$ is defined as the minimal value of black hole mass for which the semiclassical extrapolation can be trusted. Typically one expects that the construction of Eardley and Giddings holds for $M_{BH} \gg M_D$ and a semiclassical black hole will only form if, e.g., $M_{BH}\ge 3 M_D$.  For the numerical estimates  CTEQ5 PDFs were used for which an unofficial mathematica version is available on the webpage of the CTEQ collaboration.  $Q\sim M_D$ was assumed. The functions $y(z)$ were fitted to the curves given in \cite{Yoshino:2002tx}. Note that there could be a suppression of the quantum cross-section relative to the extrapolated semi-classical one, i.e., the cross section is reduced dramatically as the black hole mass drops below $\sim 5 M_D$, but this would require the existence of some small dimensionless parameters characterizing strong gravitational scattering. It was assumed otherwise.

QBHs are not expected to have high angular momentum. The incoming partons are effectively objects which are extended in space-time, their typical size is fixed by $M_D^{-1}$ (i.e., due to a minimal length imposed by quantum gravity \cite{Calmet:2004mp}), which is also the interaction range of the semiclassical formation process in the limit of a quantum black hole. Thus, the impact parameter and hence the angular momentum of the QBH are small -- at impact parameter $M_D^{-1}$ the classical angular momentum would be order one at most. A classical black hole of this size with large angular momentum would have to spin at faster than the speed of light. Thus, the spin down process before the final explosion discussed in the context of semi-classical black holes does not take place here. QBHs decay immediately to a small number of final states. 

Generically speaking, QBHs form representations of SU(3)$_c$ and carry a QED charge. The process of two partons $p_i$, $p_j$ forming a quantum black hole in the $c$ representation of SU(3)$_c$ and charge $q$ as: $p_i+p_j \to$ QBH$_c^q$ is considered in  \cite{Calmet:2008dg}. The following different transitions are possible at a proton collider:
\begin{itemize}
\item[a)] ${\bf 3} \times {\bf \overline 3}= {\bf 8} + {\bf 1}$ \\
\item[b)] ${\bf 3} \times {\bf 3}= {\bf 6} + {\bf \overline 3}$\\
\item[c)] ${\bf 3} \times {\bf 8}= {\bf 3} + {\bf \overline 6}+ {\bf 15}$\\
\item[d)] ${\bf 8} \times {\bf 8}= {\bf 1}_S + {\bf 8}_S+ {\bf 8}_A+{\bf 10} + {\bf \overline{10}}_A+ {\bf 27}_S$
\end{itemize}
Most of the time the black holes which are created carry a SU(3)$_c$ charge and come in different representations of SU(3)$_c$. This has important consequences for the production of QBHs. For example the production cross-section of a QBH$_1^0$ is given by
\begin{eqnarray}
\sigma^{pp}(s,x_{min},n,M_D) &=& \int_0^1 2z dz \int_{\frac{(x_{min} M_D)^2}{y(z)^2 s}}^1 du \int_u^1 \frac{dv}{v}  \\ \nonumber && \times F(n) \pi r_s^2(us,n,M_D)
\\ \nonumber &&
\left( \frac{1}{9} \sum_{i,j=q,{\bar q}} f_i(v,Q) f_{\bar j}(u/v,Q)
+\frac{1}{64}  f_g(v,Q) f_g(u/v,Q) \right)
\end{eqnarray}
where $i,j$ runs over all the quarks and anti-quarks subject to the constraint of QED charge neutrality,
and $f_q, f_g$ are the quark and gluon PDFs. For the production of a specific member (i.e., with specified color) of the octet QBH$_8^0$, one finds the same expression.

Since the total cross-section is known, at least semi-classically, it is straightforward to estimate the decay width in the same spirit of extrapolation. It is given  for the four dimensional model by:
\begin{eqnarray}
\Gamma({\rm QBH}_c^q \to p_1... p_f)\sim \left( 2 \pi  \left(\frac{1}{(2 \pi)^2}\right)^{(n_f-1)} \left(\frac{1}{2}\right)^{(n_f-1)} \right) \pi r_s^2 M_{BH}^3~~.
\end{eqnarray}
For quantum black holes one expects that the number of particles  $n_f$ in the final state is small, e.g., two or three. The three particle final state is strongly suppressed with respect to the two particle final state due to phase space. One thus typically has, in four dimensions,
\begin{eqnarray}
\Gamma = \frac{1}{4 \pi} \frac{M_{BH}^5}{M_D^4}
\label{width}
\end{eqnarray}
which for quantum black holes is of the order of $M_{BH}/4\pi \sim 80$ GeV for a quantum black hole with a mass of one TeV. Although consistent with the assumptions stated above, the factor of $4 \pi$ in this width could be larger in reality; for example, the sum over multiple decay channels is neglected. The actual decay width is model dependent. Another argument in favor of the decay of a quantum black hole to a two particle final state is that if the center of mass collision energy is lowered it should match the $2 \to 2$ cross-section with an exchange of a graviton. In studies of gravitational scattering by Amati, Ciafaloni and Veneziano \cite{Amati:2007ak}, evidence for an absorptive part of the forward amplitude is found near what would be the threshold of black hole formation using the Hoop Conjecture \cite{hoop}. This is consistent with the picture of quantum black holes presented above as being gravitational bound states of, e.g., two particles.

A QCD-singlet quantum black hole which is also neutral under U(1)$_{em}$, denoted as QBH$_1^0$, will decay to any combination of Higgs boson, leptons, quarks as well as gauge bosons and gravitons, e.g. QBH$_1^0 \to e^+ +e^-$, QBH$_1^0 \to e^+ + \mu^-$, QBH$_1^0 \to q_i +\bar q_i$ etc., as long as the global final state is neutral under QCD and U(1)$_{em}$. Because of the number of colored fermions in the standard model, most of the time the QBH$_1^0$ will decay to two jets. 

An octet black hole which is U(1)$_{em}$ neutral, QBH$^0_8$, can decay to a quark and an anti-quark of the opposite charge or to a gluon and a neutral particle such as a Z-boson or a photon. Further analysis of this channel depends on whether one imposes that there must exist a Lorentz invariant effective field theory description of black hole reactions. If one assumes that Lorentz invariance is not violated, then transitions of the type $q_i +g \to$ QBH$_c^q \to q_k +q_j$, where $q_i$ are quarks and $g$ is a gluon, will not take place as it is impossible to write down a Lorentz invariant local operator linking three fermions and a spin one gauge boson. 

A U(1)$_{em}$ charged triplet black hole QBH$_3^q$ can decay to a quark of charge $q$ and a gluon. These black holes will have two jets in the final state. Other decay modes of the QBH$_3^q$ which violate Lorentz conservation are for example quark+photon, quark+Z-boson, quark+graviton, quark+neutrino and quark+anti-neutrino. Similar considerations apply to black holes in higher representations:  QBH$_{10}^0$, QBH$_{\overline{10}}^0$,
QBH$_{27}^0$, QBH$_{\overline{6}}^q$, QBH$_{15}^q$,  QBH$_{6}^q$ and  QBH$_{\overline{3}}^q$.

The discussion thus far has been fairly model independent, with the exception of the production cross-section for QBHs which in the case of the model of Randall and Sundrum receives a further suppression due to the warping of the extra-dimension \cite{Meade:2007sz}.  

\subsubsection{Cross-sections at the LHC}

The inclusive production rate at the LHC of quantum black holes have been calculated in \cite{Calmet:2008dg}. The results are given in table 
(\ref{table1QBH}).  As expected, quantum black hole processes dominate over semiclassical black holes. In ADD  and in the large hidden sector model (LHS) it was assumed in \cite{Calmet:2008dg} that semiclassical black holes form already for $x_{min}=3$ whereas in RS we took $x_{min}=5$. As discussed above these assumptions are very optimistic. In the four-dimensional case, because of the large hidden sector, a non-negligible fraction of the quantum black holes (i.e. the neutral ones) decay invisibly into that hidden sector. In ADD some missing energy will be emitted in the bulk via graviton decay of QBHs. In RS, because of the mass gap, most of the energy goes in the brane and QBHs thus decay visibly.

\begin{table}[tb] \label{table1QBH}
\resizebox{\textwidth}{!}{
%\begin{center} 
\begin{tabular}{|c|c|c|c|}
\hline
models & $\sigma$(p+p $\to$ any QBH)  in fb &$\sigma$(p+p $\to$ sc-BHs) in fb& 
  $\sigma$(p+p $\to$ m.e.) in fb \\
\hline
RS& $1.9 \times 10^6$& $ 151$ & $\sim$ none\\
\hline
ADD $n=5$& $9.5 \times 10^6$& $3.1 \times 10^4$& some\\
\hline
ADD $n=6$& $1.0 \times 10^7$& $3.2 \times 10^4$ &some\\
\hline
ADD $n=7$& $1.1 \times 10^7$& $2.9 \times 10^4$ & some\\
\hline
LHS  & $1 \times 10^5$  & $5 \times 10^3$ & $744$\\
\hline
\end{tabular}}
%\end{center}
\caption{Cross-sections for the production of quantum black holes and semiclassical (sc) black holes. The missing energy  (m.e.) component is also indicated. We take the reduced Planck mass to be 1 TeV and thus restrict our considerations to ADD with $n\ge 5$ since lower dimensional models with $M_D=1$ TeV are ruled out by astrophysical data. Note that the bound on the reduced Planck mass in four-dimensions is only of the order of 488 GeV \cite{Calmet:2008rv}.}
\end{table}

Due to conservation of color, most quantum black hole events at the LHC will give rise to two jets. However, the standard model background can be large. Two interesting signatures with less or no background are: proton+proton $\to$ QBH $\to$ lepton + anti-lepton of another generation and proton+proton $\to$ QBH $\to$ lepton + jet. The latter can only occur if one allows violation of baryon and lepton number (which is model dependent; for example these symmetries might be gauged). For example, consider p+p $\to$ QBH$^{-2/3}_{\bar{3}} \to l^- + \bar d$. If described by a local effective field theory, it could correspond to the operator ${\cal O} = \bar{u^c}_L d_L \bar{e}_L d_R$. The reaction with an anti-lepton in the final state can be mediated by $qqql$. Processes like p+p $\to$ QBH$^{1/3}_{\bar{3}} \to \gamma + \bar d$ would correspond to an operator connecting three fermions to a vector particle, which violates Lorentz invariance.

Since gravity is democratic (i.e., it couples equally to all flavors) it is expected  that
$\sigma$(p+p $\to$ QBH $\to e$ + jet) $~=~ \sigma$(p+p $\to$ QBH $\to \mu$ + jet)$~=~\sigma$(p+p $\to$ QBH $\tau$+ jet), neglecting the masses of the fermions. Some cross-sections  with a lepton and a jet in the final state are listed in  table (\ref{table2QBH}). The final state lepton can belong to the first, second or third generation. Some QBH$_{\bar 3}$ black holes will lead to a remarkable signature with a jet and a lepton back to back with high $p_T$. The lepton can be a neutrino, in which case the signature is missing energy with a high $p_T$ jet. Note that gauge bosons and the Higgs boson can appear in the final state: e.g., QBHs can decay to a Z and a jet or a Higgs boson + jet. The cross-section $\sigma$(p+p $\to$ Z+jet)  is equal to $3/2 \times \sigma$(p+p $\to \gamma$+jet). One can also have final states  involving missing energy, e.g., QBH $\to$ graviton+jet, $\nu$+jet and  $\bar \nu$+jet. Other interesting decay modes of QBHs involve a gluon and a photon in the final state. These can also appear in string theory as recently pointed out by Anchordoqui et al. \cite{Anchordoqui:2007da,Anchordoqui:2008hi}. Note that the width obtained in the calculation for decay of lowest massive Regge excitations of open strings \cite{Anchordoqui:2008hi} agrees with our result in eq. ({\ref{width}). As discussed in  \cite{Anchordoqui:2007da,Anchordoqui:2008hi}, the Lorentz conserving transitions $q+g \to$ QBH $\to \ q+ \gamma$ or 
$g+g \to$ QBH $\to \ g+ \gamma$ could lead to interesting signals, although the standard model background might be larger in these cases.

\begin{table}[tb]\label{table2QBH}
\resizebox{\textwidth}{!}{
%\begin{center}
\begin{tabular}{|c|c|c|c|c|c|}
\hline
cross-sections in fb & LHS & RS& ADD $n=5$ & ADD $n=6$ & ADD $n=7$   \\
\hline
$\sigma$(p+p $\to$ QBH$^{4/3}_{\bar{3}} \to l^+ + \bar d )$ &  372&$5.8\times10^3$ &$3.3\times 10^4$ &$3.7\times 10^4$ &$4\times 10^4$ \\
\hline
$\sigma$(p+p $\to$ QBH$^{-2/3}_{\bar{3}} \to l^- + \bar d )$& 47& $734$&$3.7\times 10^3$ &$4\times 10^3$ & $4.2\times 10^3$
\\
\hline
 $\sigma$(p+p $\to$ QBH$^{1/3}_{\bar{3}} \to \nu_i + \bar d )$& 160& $2.5\times10^3$& $1.4\times 10^4$& $1.5\times 10^4$&$1.6\times 10^4$\\
 \hline
  $\sigma$(p+p $\to$ QBH$^{-2/3}_{\bar{3}} \to \nu_i + \bar u )$& 47&$734$ & $3.7\times 10^3$ &$4\times 10^3$ & $4.2\times 10^3$  \\
\hline
$\sigma$(p+p $\to$ QBH$^{-2/3}_{\bar{3}} \to \gamma+ \bar u )$& 47&$734$ & $3.7\times 10^3$ &$4\times 10^3$ & $4.2\times 10^3$  \\
\hline
$\sigma$(p+p $\to$ QBH$^{1/3}_{\bar{3}} \to \gamma + \bar d )$& 160& $2.5\times10^3$& $1.4\times 10^4$& $1.5\times 10^4$&$1.6\times 10^4$\\
 \hline
$\sigma$(p+p $\to$ QBH$^0_1$ $\to e^+ + \mu^-)$& 0& 93&$447$ &$491$& $511$ \\
\hline
\end{tabular}}
%\end{center}
\caption{Some possible final states in quantum black hole decay for the models LHS, RS and ADD. Gravity is democratic, one thus expects the same cross-sections for final states with any charged lepton combination. Note that if the neutrino is a Majorana particle the cross-section $\sigma$(p+p $\to$ QBH$^{-2/3}_{\bar{3}} \to \nu_i + \bar u )$ is $11/9$ times larger than what is given in the table, since one cannot differentiate $\nu$ from $\bar \nu$. If the neutrino is a Dirac type particle, then one has $\sigma$(p+p $\to$ QBH$^{-2/3}_{\bar{3}} \to \nu_i + \bar u )$ = $\sigma$(p+p $\to$ QBH$^{-2/3}_{\bar{3}} \to \bar \nu_i + \bar u )$. Note that the sum over the polarization of the photon for the cross-sections $\sigma$(p+p $\to \gamma$+jet) has been performed. The cross-section $\sigma$(p+p $\to$ Z+jet) = $3/2 \times \sigma$(p+p $\to \gamma$+jet).}
\end{table}

Another interesting signature of quantum black holes are decays of neutral, SU(3)$_c$ singlet holes to two leptons of different generations with opposite charge. These decays are highly suppressed in the LHS model since QBH$_1^0$ will decay invisibly in the hidden sector. However, in ADD and RS these signatures would be characteristic of quantum black holes.  The cross-sections $\sigma$(p+p $\to$ (neutral QBHs) $\to e^+ +\mu^-)$ can be found in table (\ref{table2QBH}). Again because of the universality of gravity one expects: $\sigma$(p+p $\to$ (neutral QBHs) $\to e^++\mu^-) ~=~ \sigma$(p+p $\to$ (neutral QBHs) $\to e^+ + \tau^-) ~=~ \sigma$(p+p $\to$ (neutral QBHs) $\to \tau^+ + \mu^-) ~=~ \sigma$(p+p $\to$ (neutral QBHs) $\to l^+ + l^-)$ where the l$^\pm$ can be any lepton.

Clearly a more thorough analysis for the LHC needs to be performed. In particular the  event generators which have been written for semi-classical black holes   \cite{Dai:2007ki,Cavaglia:2006uk,DL,Harris:2003db} need to be adapted to take into account the hidden sector scenarios and in some cases the quantum black holes cases.

\section{Conclusion}
 
 The realization that the scale of quantum gravity could be much lower than naively expected using dimensional analysis and potentially in the TeV region has triggered fascinating developments not only in theoretical particles physics but also in relativity and astroparticle physics and cosmology. On the relativity side, these models were a motivation to study the fundamental question of when a small black hole is formed in the collision of two particles with a non zero impact parameter. This problem is now well understood and has deep implications for searches of quantum black holes at colliders. It is very unlikely that semi-classical black holes will be produced at the LHC because the energy available is too low. However, plenty of small non-thermal quantum black holes could be produced.  If small black holes are produced, it would be a fascinating opportunity to probe the symmetries of quantum gravity and learn how to unify quantum mechanics and General Relativity.
 
 On the field theoretical side, precise calculations of the emission of massive or massless gravitons have been performed using linearized General Relativity. A more recent development is that models with low scale quantum gravity typically have problems with unitarity below the scale at which gravity becomes strong. An important implication is that if a model belonging to this class was relevant to nature, the first signal of physics beyond the standard model would not be of gravitational nature, but rather a sign of a non local interaction, for example, a stringy interaction.
 
 A major effort has been done to probe these ideas  using  astroparticle physics data as well as  cosmological ones. The opinion of the author is that not all paths have yet been studied and more exciting results are to be expected from collaborations between theoretical physicists and astronomers.

 With the Large Hadron Collider starting to collect data, we could be on the verge of discovering extra-dimensions or a large hidden sector. It is however clear that the first signals of these models would not be linked to gravitational physics but rather to the mechanism that restores unitarity below the reduced Planck mass. This is clearly an exciting time for physics.

\section*{Acknowledgments}

I would like to thank Michael Atkins, Priscila de Aquino, Stephen Hsu, Wei Gong, David Reeb and Thomas Rizzo for very enjoyable collaborations on topics related to this review. This work in supported in part by the European Cooperation in Science and Technology (COST) action MP0905 "Black Holes in a Violent Universe".

\section*{Appendix A: Feynman rules}

In this appendix the necessary Feynman rules are summarized. The conventions are those of \cite{Han:1998sg}:
\begin{align}
&C_{\mu\nu,\rho\sigma} & =\quad & \eta_{\mu\rho} \eta_{\nu\sigma} + \eta_{\mu\sigma} \eta_{\nu\rho} -\eta_{\mu\nu} \eta_{\rho\sigma}\\ 
\nonumber \\ 
&D_{\mu\nu,\rho\sigma}(k_1,k_2) & =\quad & \eta_{\mu\nu} k_{1\sigma} k_{2\rho} - 
\eta_{\mu\sigma} k_{1\nu} k_{2\rho}- \eta_{\mu\rho} k_{1\sigma} k_{2\nu} + \eta_{\rho\sigma} k_{1\mu} k_{2\nu}   \\ \nonumber &&&   - 
\eta_{\nu\sigma} k_{1\mu} k_{2\rho}- \eta_{\nu\rho} k_{1\sigma} k_{2\mu} + \eta_{\rho\sigma} k_{1\nu} k_{2\mu}  \\ \nonumber
\\ 
&E_{\mu\nu,\rho\sigma}(k_1,k_2) & =\quad & \eta_{\mu\nu} ( k_{1\rho} k_{1\sigma} + k_{2\rho} k_{2\sigma}+k_{1\rho} k_{2\sigma}) 
\\ \nonumber &&&   
-\eta_{\nu\sigma} k_{1\mu}k_{1\rho} -\eta_{\nu\rho} k_{2\mu}k_{2\sigma} 
 -\eta_{\mu\sigma} k_{1\nu}k_{1\rho} -\eta_{\mu\rho} k_{2\nu}k_{2\sigma} 
\end{align}
The propagator for the quarks and gluons are respectively given by
\begin{eqnarray}
\frac{i(\fmslash p +m)}{p^2-m^2+i \epsilon}
\end{eqnarray}
and
\begin{eqnarray}
\frac{-i \delta^{ab}}{k^2} \left (g^{\mu\nu}- \frac{k^\mu k^\nu}{k^2} (1-\xi)\right).
\end{eqnarray}

The vertices describing the interactions of the graviton are given by

%\resizebox{\textwidth}{!}{
\begin{tabular}{c>{\raggedright}m{5in}}
 & 
 \tabularnewline
\includegraphics[scale=0.18]{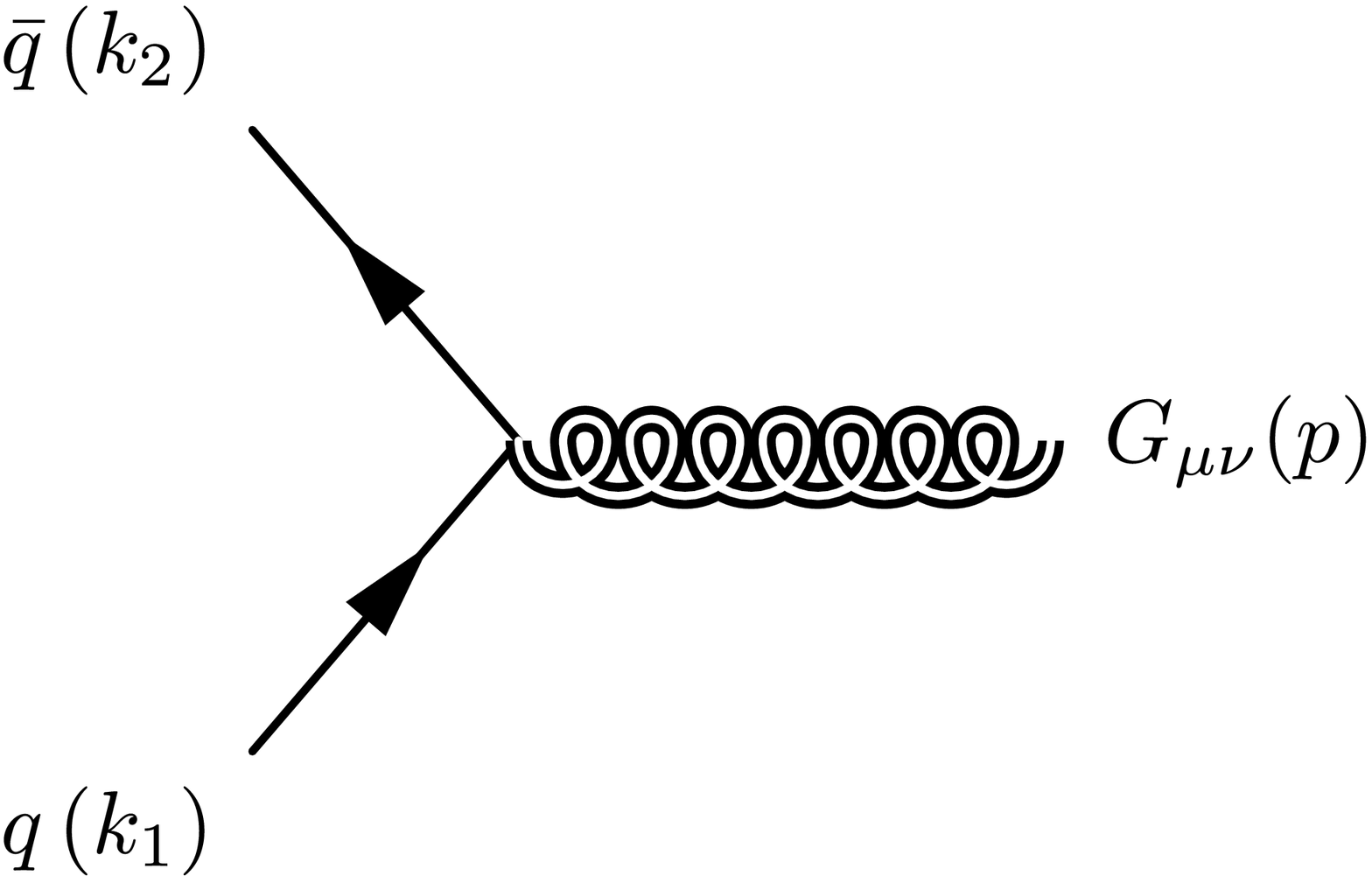} & $=-i\,\frac{\kappa}{8}\left\{ \gamma_{\mu}\left(k_{1\,\nu}+k_{2\,\nu}\right)+\gamma_{\nu}\left(k_{1\,\mu}+k_{2\,\mu}\right)
-2\eta_{\mu\nu}\left(\not k_{1}+\not k_{2}\right)\right\} $
$\qquad$$\qquad$\vspace{1in}
\tabularnewline
 & \tabularnewline
\includegraphics[scale=0.18]{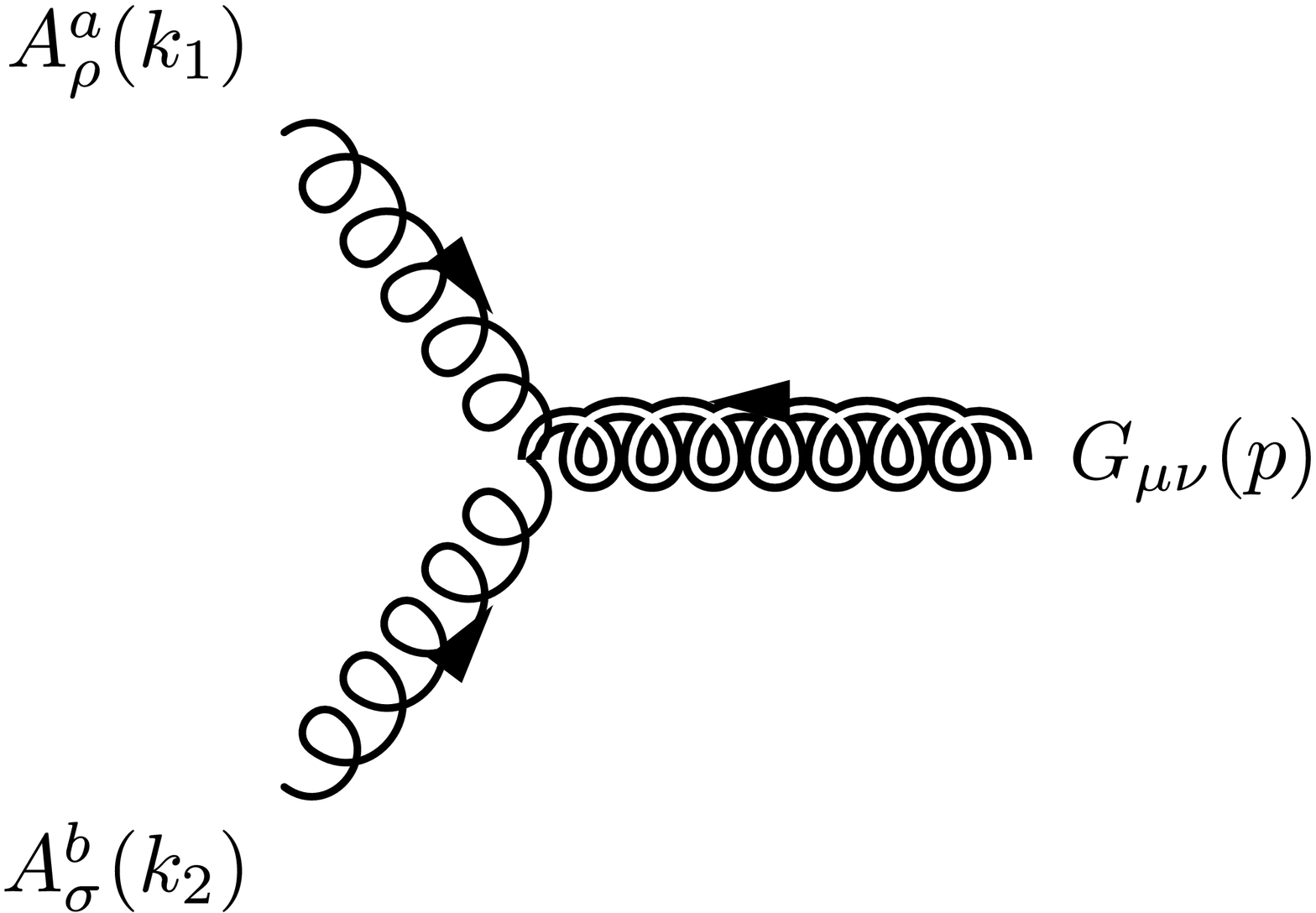} & $=i\,\frac{\kappa}{2}\delta^{ab}\left\{ \left[k_{1}\cdot k_{2}\right]C_{\mu\nu\rho\sigma}+D_{\mu\nu\rho\sigma}\left(k_{1},k_{2}\right)+\xi^{-1}E_{\mu\nu\rho\sigma}\left(k_{1},k_{2}\right)\right\} $
$\qquad$$\qquad$\vspace{1in}
\tabularnewline
& \tabularnewline
\includegraphics[scale=0.18]{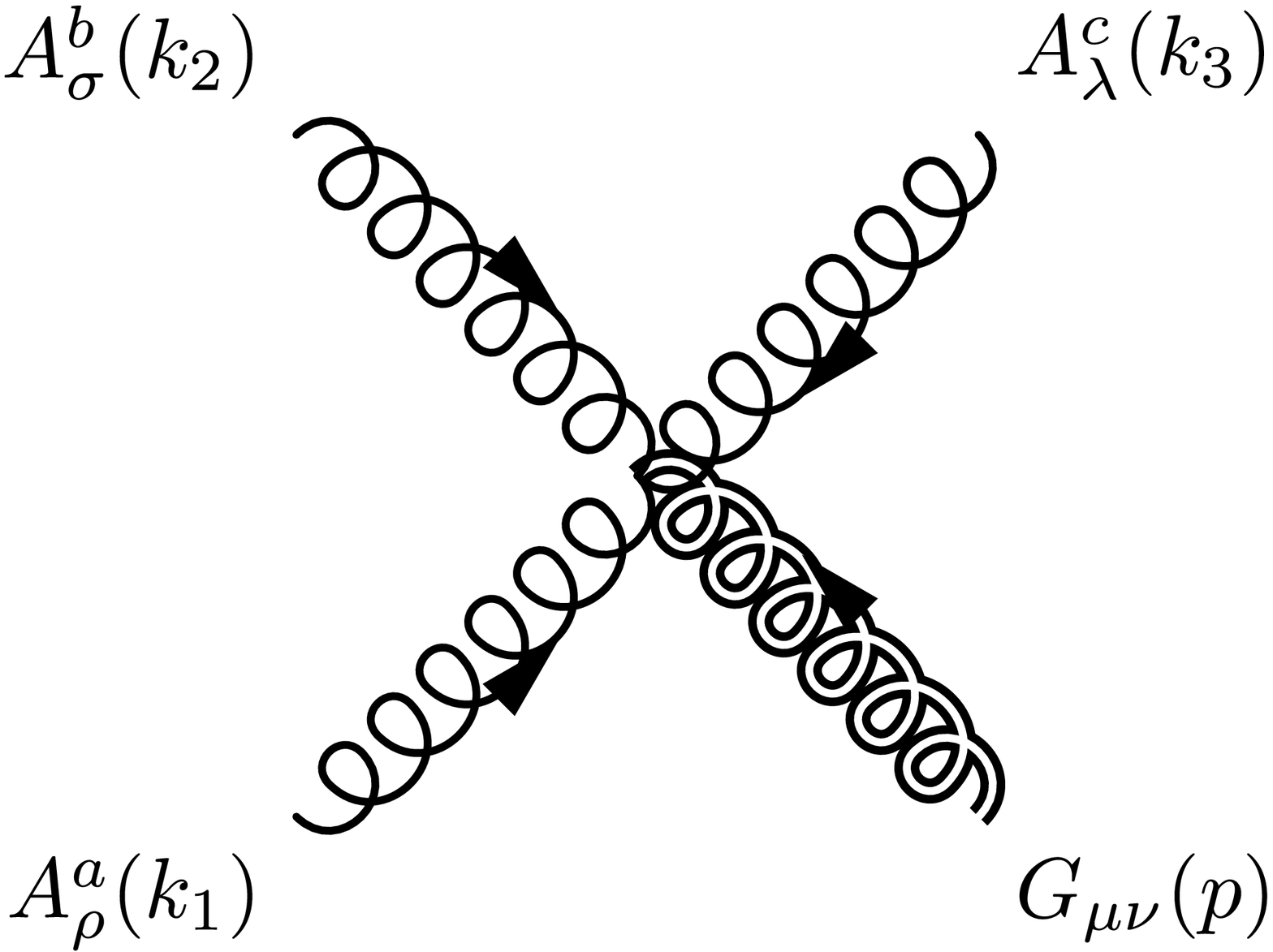} & $=g\,\frac{\kappa}{2}f^{abc}\left\{ C_{\mu\nu\rho\sigma}\left[k_{1\,\lambda}-k_{2\,\lambda}\right]+C_{\mu\nu\rho\lambda}\left[k_{3\,\sigma}-k_{1\,\sigma}\right]\right.$

\medskip{}

$\qquad\qquad\left.+C_{\mu\nu\sigma\lambda}\left[k_{2\,\rho}-k_{3\,\rho}\right]+F_{\mu\nu\rho\sigma\lambda}\left(k_{1},\, k_{2},\, k_{3})\right)\right\} $
$\qquad$$\qquad$\vspace{1in}
\tabularnewline
\end{tabular}

\begin{tabular}{c>{\raggedright}m{5in}}
 \tabularnewline
\includegraphics[scale=0.18]{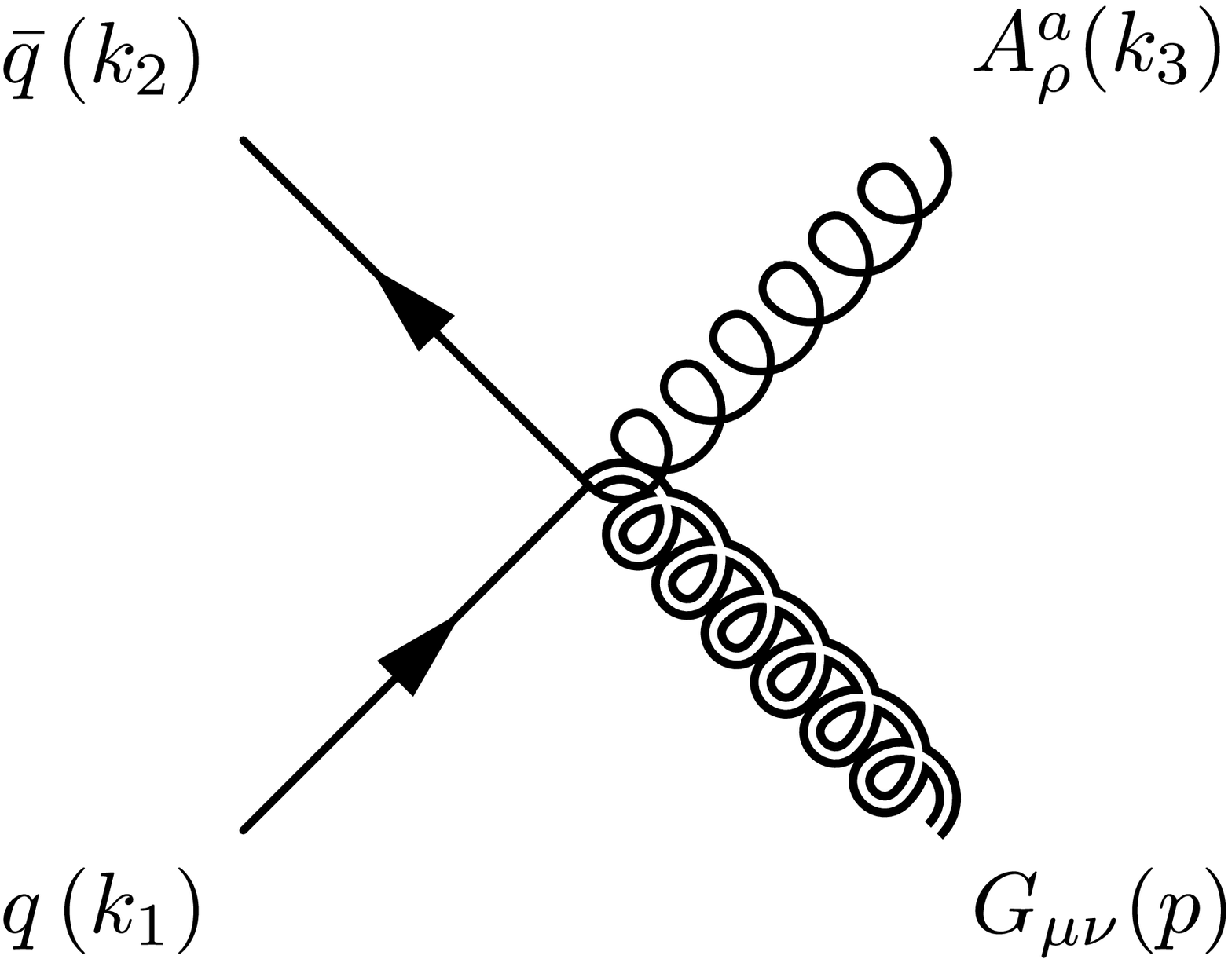} & $=i\, g\,\frac{\kappa}{4}\, T^{a}\left\{ C_{\mu\nu\rho\sigma}-\eta_{\mu\nu}\eta_{\rho\sigma}\right\} \gamma^{\sigma}$
$\qquad$$\qquad$\vspace{1in}
\tabularnewline
 & \tabularnewline
\end{tabular}
%}
with the understanding that $\kappa= 16 \pi G_N$ is scale dependent. $\xi$ is a gauge fixing parameter not to be confused with the non-minimal coupling.
%\pagebreak{}

Finally, we also make use of the standard model three particles vertices:

\begin{center}
\textbf{\textsc{\LARGE }}\begin{tabular}{c>{\raggedright}m{5in}}
 & \tabularnewline
\includegraphics[scale=0.18]{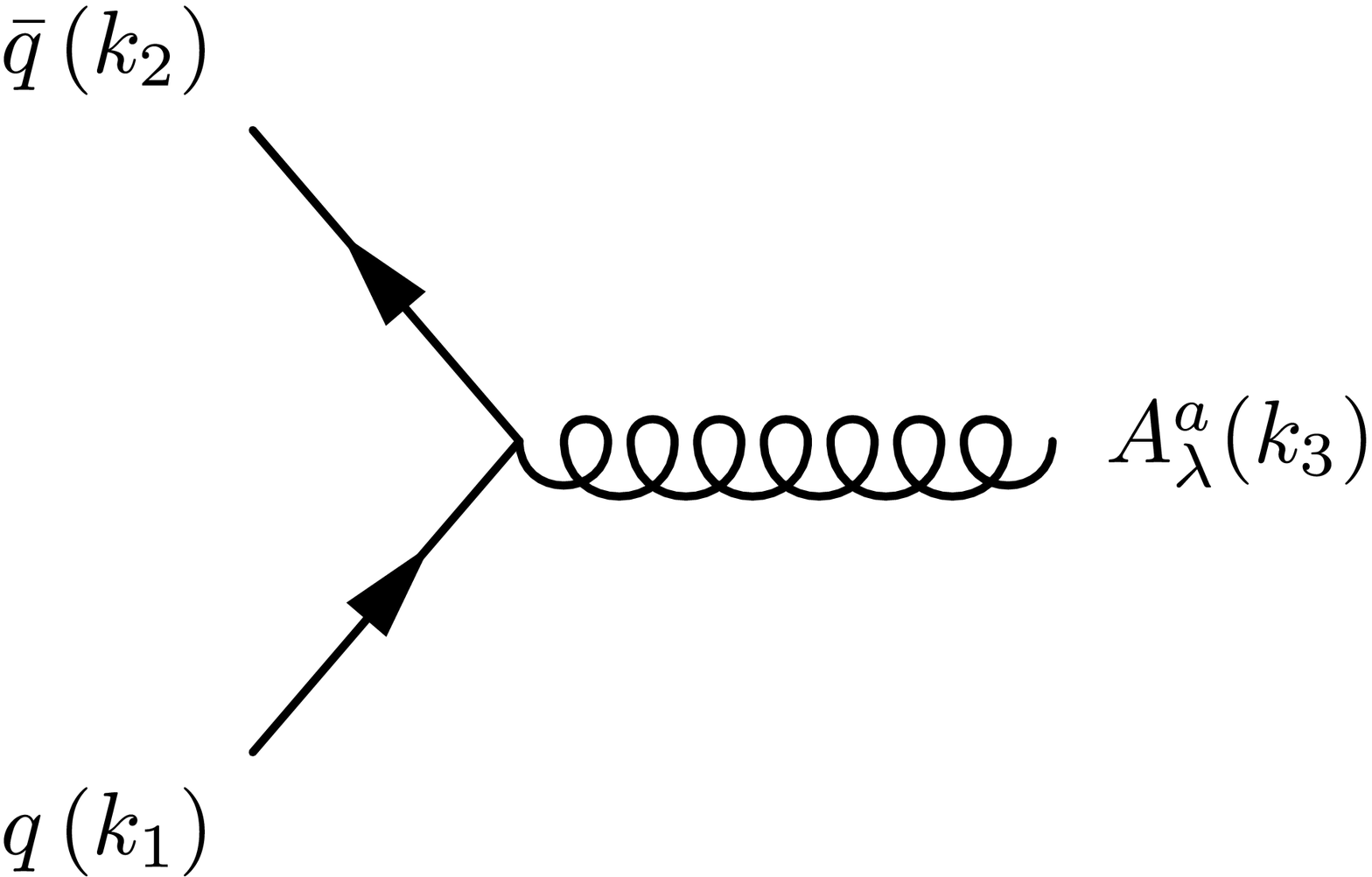} & $=i\, g\, T^{a}\,\gamma^{\lambda}$ $\qquad$$\qquad$$\qquad$\vspace{1in}
\tabularnewline
 & $ $\tabularnewline
\includegraphics[scale=0.18]{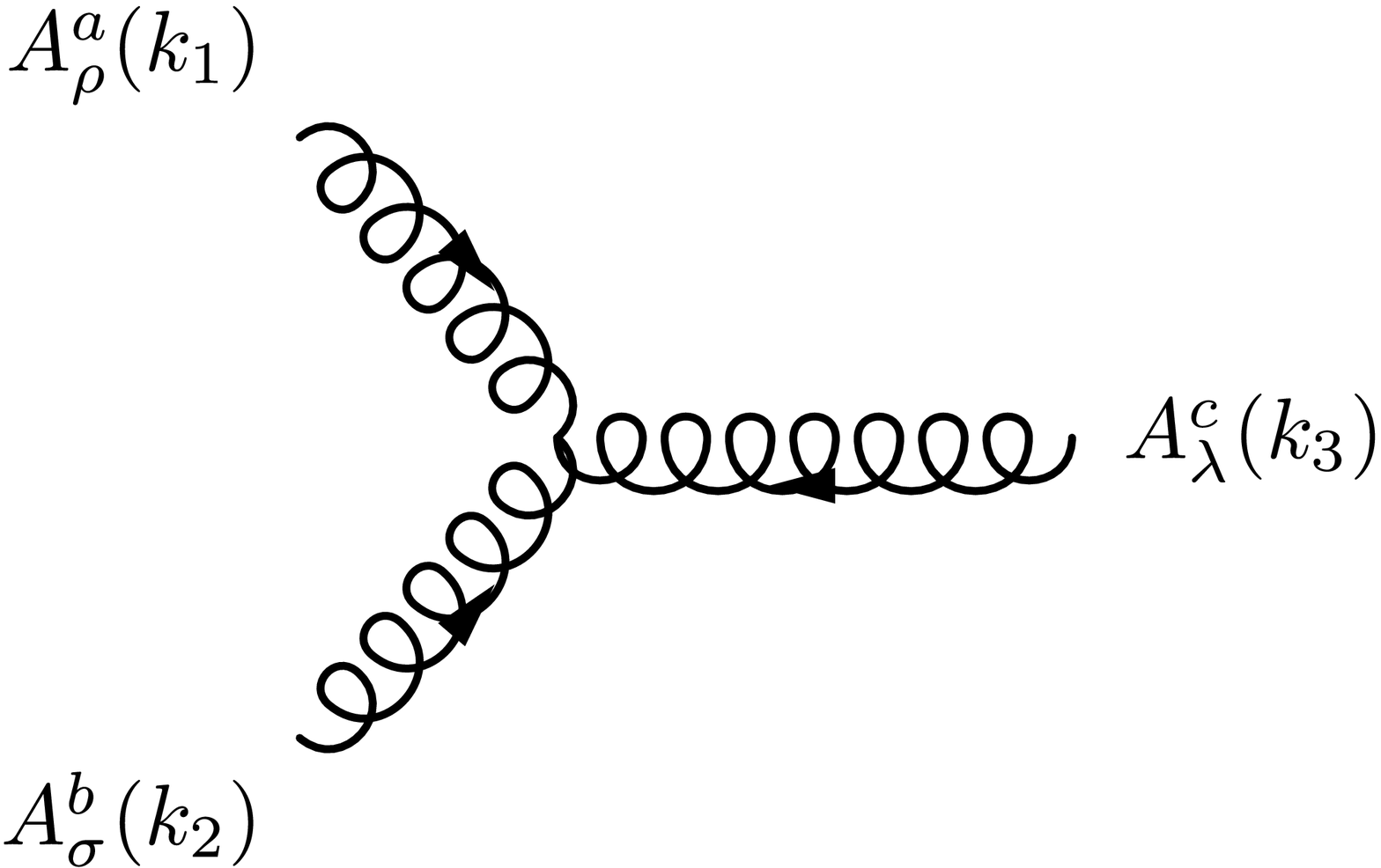} & $=g\, f^{abc}\left\{ \eta_{\rho\sigma}\left(k_{1\,\lambda}-k_{2\,\lambda}\right)+\eta_{\sigma\lambda}\left(k_{2\,\rho}-k_{3\,\rho}\right)+\eta_{\lambda\rho}\left(k_{3\,\sigma}-k_{1\,\sigma}\right)\right\} $
$\qquad$$\qquad$\vspace{1in}
\tabularnewline
 & \tabularnewline
\end{tabular}
\par\end{center}

\section*{Appendix B: Renormalization of Newton's constant}

Consider the contribution of a scalar field minimally coupled to gravity. We follow the presentation of Larsen and Wilczek \cite{Larsen:1995ax} (see also \cite{DeWitt,BirrellDavies}). 
The one-loop effective action $W$ is defined through
\begin{eqnarray}
e^{-W}&=&\int {\cal D}\phi~e^{-\frac{1}{8\pi}\int\phi (-\Delta+m^2)\phi} \\ 
\nonumber
&=&[{\rm det}(-\Delta+m^2)]^{-\frac{1}{2}}~.
\end{eqnarray}
We define the heat kernel
\begin{eqnarray}
H(\tau) \equiv {\rm Tr}e^{-\tau\Lambda}=\sum_i e^{-\tau\lambda_i}~,
\end{eqnarray}
where $\lambda_i$ are the eigenvalues of $\Lambda=-\Delta + m^2$.
Then the effective action reads
\begin{eqnarray}
\label{W}
W=\frac{1}{2}{\rm ln}\,{\rm det}\Lambda=\frac{1}{2}\sum_i {\rm ln}\lambda_i
=-\frac{1}{2}\int_{\epsilon^2}^\infty
d\tau \frac{H(\tau)}{\tau}~.
\end{eqnarray}
The integral over $\tau$ is divergent and has to be regulated by an ultraviolet cutoff $\epsilon^2$.
The heat kernel method can be used to regularize the leading divergence of this integral. This technique does not violate general coordinate invariance. One can write
\begin{eqnarray}
H(\tau)=\int dx~G(x,x,\tau)~,
\end{eqnarray}
where
the Green's function $G(x,x^{\prime},\tau)$ satisfies the differential equation
\begin{eqnarray}
(\frac{\partial}{\partial\tau}-\Delta_x)G(x,x^{\prime},\tau)&=&0~; \\
G(x,x^{\prime},0)&=&\delta(x-x^\prime)~.
\end{eqnarray}
In flat space one has
\begin{eqnarray}
G_0(x,x^\prime,\tau) = \left(\frac{1}{4\pi\tau}\right)^{2}
\exp{\left (-\frac{1}{4\tau}(x-x^\prime)^2 \right )}~,
\end{eqnarray}
but in general one must express the covariant Laplacian in local coordinates
and expand for small curvatures.

The result is \cite{BalianBloch}
\begin{eqnarray} \label{BB}
H(\tau)&=& \frac{1}{(4\pi\tau)^2}    
\Big( \int d^4x\, \sqrt{-g}
\\ \nonumber &&~+~ \frac{\tau}{6} \int d^4x \, \sqrt{-g} \, R   
~+~ {\cal O}(\tau^\frac{3}{ 2})\Big)~.
\end{eqnarray}
Plugging this back into (\ref{W}), one obtains the renormalized Newton constant
\begin{eqnarray} \label{cutdependence}
\frac{1}{G_{N, \rm ren}} ~=~
 \frac{1}{G_{N, \rm bare} } + \frac{1}{12 \pi \epsilon^2}~,
\end{eqnarray}
so that $G_{\rm N, ren}$, relevant for long-distance measurements, is much smaller than the bare value if the scalar field is integrated out ($\epsilon \rightarrow 0$).

Up to this point our results have been in terms of old-fashioned renormalization: we give a relation between the physical observable $G_{N, \rm ren}$ and the bare coupling $G_{N, \rm bare}$. A modern Wilsonian effective theory would describe modes with momenta $\vert k \vert < \mu$. Modes with $\vert k \vert > \mu$ have been integrated out and their virtual effects already absorbed in effective couplings $g( \mu )$. In this language, $G_{N, \rm ren} = G _N( \mu = 0 )$ is appropriate for astrophysical and other long-distance measurements of the strength of gravity.

A Wilsonian Newton constant $G_N( \mu )$ can be calculated via a modified version of the previous method, this time with an infrared cutoff $\mu$. For example, (\ref{W}) is modified to
\begin{eqnarray}
\label{W2}
W=-\frac{1}{2}\int_{\epsilon^2}^{\mu^{-2}}
d\tau \frac{H(\tau)}{\tau}~.
\end{eqnarray}
The resulting Wilsonian running of Newton's constant is
\begin{eqnarray}
\frac{1}{G _{N}( \mu )} ~=~
\frac{1}{G_{N}(0) } - \frac{\mu^2}{12 \pi}~,
\end{eqnarray}
 or
 \begin{eqnarray}
 \label{Nrunning}
\frac{1}{G_{N} ( \mu )} ~=~
\frac{1}{G_{N}(0) } - N \frac{\mu^2}{12 \pi} 
\end{eqnarray}
for $N$ scalars or Weyl fermions, as can be shown by a similar functional calculation. Cf.~\cite{Larsen:1995ax}, who also derive the opposite sign in the gauge boson case.

We note that (\ref{cutdependence}) and (\ref{Nrunning}) are only valid to leading order in perturbation theory. As we near the scale of strong quantum gravity, $\mu_*$, we lose control of the model. However, it seems implausible that the sign of the beta function for Newton's constant will reverse, so the qualitative prediction of weaker gravity at low energies should still hold. 

\bigskip
\bigskip

%\section{References}

%\begin{thebibliography}{000} %for 3 digits
%\begin{thebibliography}{00}  %for 2 digits

\end{document}